\documentclass[aps,pra,twocolumn, superscriptaddress,showpacs]{revtex4-1}%
\usepackage{amsfonts}
\usepackage{amsmath}
\usepackage{amssymb}
\usepackage{graphicx}%
\usepackage[english]{babel}
\usepackage{hyperref}

\usepackage[usenames,dvipsnames]{color}
\makeatletter
\@ifundefined{textcolor}{}
{%
 \definecolor{BLACK}{gray}{0}
 \definecolor{WHITE}{gray}{1}
 \definecolor{RED}{rgb}{1,0,0}
 \definecolor{GREEN}{rgb}{0,1,0}
 \definecolor{BLUE}{rgb}{0,0,1}
 \definecolor{CYAN}{cmyk}{1,0,0,0}
 \definecolor{MAGENTA}{cmyk}{0,1,0,0}
 \definecolor{YELLOW}{cmyk}{0,0,1,0}
 }

\newcommand{\beq}{\begin{equation}}
\newcommand{\eeq}{\end{equation}}
\newcommand{\nn}{\nonumber}
\newcommand{\ket}[1]{|#1\rangle}
\newcommand{\bra}[1]{\langle #1|}

\usepackage{babel}
\begin{document}

\title{Quantum power boost in a nonstationary cavity-QED quantum heat engine}

\author{A. V. Dodonov
%$^{1}$
}\email{adodonov@fis.unb.br}
\affiliation{Institute of Physics and International Centre for Condensed Matter Physics,
University of Brasilia, 70910-900, Brasilia, Federal District, Brazil}

\author{D. Valente
%$^{1}$
}

\affiliation{
%$^{2,3}$
Instituto de F\'isica, Universidade Federal de Mato Grosso, 78060-900, Cuiab\'a MT, Brazil}

\affiliation{
Laboratoire Pierre Aigrain, \'Ecole Normale Sup\'erieure - Paris, Centre National de la Recherche Scientifique, 24 rue Lhomond, 75005 Paris, France}

\author{T. Werlang
%$^{1}$
}

%\singlespacing

\affiliation{
%$^{1}$
Instituto de F\'isica, Universidade Federal de Mato Grosso, 78060-900, Cuiab\'a MT, Brazil}

\begin{abstract}
We show a quantum boost in the output power of a heat engine formed by a two-level system coupled to a single-mode cavity.
The key ingredient here is the nonstationary regime achieved when some system parameter (atomic transition frequency, in our case) is subjected to a time-dependent perturbative modulation that is precisely tuned at certain frequencies.
We discuss how the extracted power can lead to amplification of the external driving field.
Quantum power boost is found both in the nonstationary Jaynes-Cummings and Rabi models, indicating that our predictions can be experimentally tested in circuit quantum electrodynamics setups.

\end{abstract}
\pacs{03.65-w, 42.50.Pq, 05.70-a}
\maketitle

%%%%%%%%%%%%%%%%%%%%%%%%%%%%%%%%%%%%%%%%%%%%%%%%%
%
\section{Introduction}
The role of quantum coherences on the power of heat engines is currently an open debate.
Ref. \cite{Kosloff2015} has shown that the output power of a quantum heat engine is a key quantum-thermodynamic signature, enabling the distinction between classical and quantum thermodynamic processes.
This brings the discussion on whether quantum coherence can be beneficial or detrimental to the power of a heat engine.
Depending on the regime under consideration, the output power can be enhanced by quantum effects, as theoretically \cite{Scully2011, Kosloff2015, Kurizki2015} and experimentally \cite{Walmsley2017} shown, or diminished \cite{Seifert2017, Pekola2016}.
The purpose of the present paper is to contribute to a better understanding of this subject in a so far unexplored regime.
Here, we theoretically show how certain \emph{nonstationary regimes} of cavity Quantum Electrodynamics (cavity-QED) can boost the power of a heat engine by means of quantum coherence.

Cavity-QED is the area that studies the interaction between the quantized electromagnetic (EM) field  and few-level emitters, e.g., cold Rydberg atoms \cite{q1}, quantum dots and wells \cite{q2,q3}, nitrogen-vacancy centers \cite{q4} and Bose-Einstein condensates \cite{q5,q6}. 
The extension of concepts of cavity-QED to superconducting circuits originated the area of circuit Quantum Electrodynamics (circuit-QED), where artificial superconducting atoms are formed by Josephson Junctions and interact with the EM field confined in microwave resonators on a chip \cite{you,nori,science,rev1,nori2017}. 

Nonstationary regimes of cavity QED englobe the scenarios in which one or several system parameters undergo a time-dependent perturbation prescribed externally \cite{jpcs,jpa}, and can be readily implemented in several circuit-QED architectures \cite{q2,q3,majer,ge,ger,ger1,v1,v2,v3,nori-n,meta,blais-exp,simmonds,schuster}.
Here we focus on weak (perturbative) harmonic modulations, whose frequencies are accurately tuned to induce quantum transitions between the system states that would be otherwise insignificant. 
The dynamical Casimir effect, consisting in the deterministic generation of quanta from the initial vacuum field state, is an emblematic example of the quantum dynamics that can be achieved in a nonstationary cavity QED regime \cite{jpcs,liberato,fujii,roberto,entangles}. 
Conversely, in the antidynamical Casimir effect (ADCE), photons can be coherently annihilated from a large class of nonvacuum initial states \cite{igor,diego,lucas,juan}. 
Besides, excitations can be deterministically transferred between the field and a far-detuned atom using the sideband transitions, as demonstrated in \cite{blais-exp,simmonds,schuster}.
Such parametric amplification and coupling mechanisms, together with the associated quantum-coherent nature of the unitary dynamics, make nonstationary cavity-QED regimes potentially useful for the implementation of thermal machines with finite output powers.
In Ref. \cite{AD2017}, for instance, we have recently demonstrated that the ADCE can be used as a resource for work extraction.

In this paper, we address the following problems.
It remained unclear whether the nonstationary cavity-QED regimes could be actually employed in a full thermodynamic cycle, due to the rather strict physical requirements that must be fulfilled.
The question whether quantum coherence would play a significant role in the output power of a given cycle was also left open.
Finally, the operational meaning of the extracted work, in terms of measurable external observables, has not been established so far. Here we investigate the above issues for a quite general cavity-QED setting and show:

(i) an Otto cycle for nonstationary cavity-QED;

(ii) a quantum boost in the output power of this cycle;

(iii) the mechanism of the driving field amplification.

This paper is organized as follows.
Sec. \ref{II} is devoted to the analysis of the dynamics of a quantum heat engine in nonstationary cavity-QED. Sec. \ref{II.A} introduces a nonstationary regime of the Jaynes-Cummings model and Sec. \ref{II.B} presents our nonstationary Otto cycle.
In Sec. \ref{II.C} we evidence that the quantum coherence lies at the root of the output power of our heat engine and in Sec. \ref{II.D} we discuss how the extracted work can amplify the modulating field.
Sec. \ref{III} shows that the quantum power boost is maintained in the nonstationary Rabi model for the ADCE and Jaynes-Cummings regimes. Finally, the conclusions are summarized in Sec. \ref{Conclusions}.

\section{Nonstationary cavity-QED quantum heat engine}
\label{II}

%%%%%%%%%%%%%%%%%%%%%%%%%%%%%%%%%%%%%%%%%%%%%%%%%%%%%%%%%%%%%%
\subsection{Nonstationary Jaynes-Cummings model}
\label{II.A}
We begin our analysis with the simplified description of nonstationary cavity-QED valid for weak atom-field couplings and \lq low\rq\ modulation frequencies (smaller than the cavity natural frequency). The nonstationary Jaynes-Cummings (JC) model \cite{jpcs,red2,Silveri} describes the interaction of a two-level atom with a single-mode cavity as
\beq\label{hjc}
H(t)/\hbar =\omega a^\dagger a +\frac{\Omega_t}{2}\sigma_z + g_t(a\sigma_+ +a^\dagger\sigma_-).
\eeq
$a$ ($a^\dagger$) is the cavity annihilation (creation) operator.
$\sigma_z = \sigma_+ \sigma_- - \sigma_- \sigma_+$, $\sigma_+=\ket{e}\bra{g}$ and $\sigma_-=\sigma_+^\dagger$ are the atomic ladder operators, with the ground (excited) state $\ket{g}(\ket{e})$.
The cavity frequency is $\omega$, the time-dependent atomic transition frequency is $\Omega_t$ and the time-dependent atom-cavity coupling strength is $g_t$ (we assume $g_t\ll\omega$).

In the nonstationary regime we assume the external modulations of the form
\beq\label{atf}
\Omega_t = \Omega_0 + \epsilon \sin(\eta t),
%chamar gt pra dizer que ele so serve pra turn-on/off
\eeq
where $\Omega_0$ is the bare atomic frequency, $\eta$ is the modulation frequency and
$\epsilon \ll \Omega_0$ is the modulation amplitude. We also assume that the atom-cavity coupling can be monotonically switched on, $g_t = 0 \rightarrow g_0$ (and off, $g_0 \rightarrow 0$).
For our protocol we assume the dispersive regime
\beq
|\Delta_-| \gg 2 g_0 \sqrt{n_{max}},
\label{disp}
\eeq
where $\Delta_- \equiv \omega-\Omega_0$ is the bare cavity-atom detuning
and $n_{max}$ is the maximum number of system excitations.
In the stationary case ($\eta = 0$) the exchange of excitations between the atom and the cavity is strongly inhibited due to the energetic mismatch $|\hbar\omega - \hbar\Omega_0| \gg \hbar g_0$. The external modulation of $\Omega_t$ can compensate for the energy mismatch, and under the resonant condition \cite{jpcs,roberto,AD2017,palermo,red1,red2}
\beq
\eta = \sqrt{\Delta_-^2+4g_0^2(n+1)} \ \approx \ |\Delta_-|
\label{nstres}
\eeq
the system exhibits complete periodic (red-sideband) transitions between the approximate states
\beq
\ket{g,n+1} \longleftrightarrow \ket{e,n},
\eeq
where $\ket{n}$ is the cavity Fock state, defined by $a^\dagger a\ket{n}=n\ket{n}$ (see details in the Appendix \ref{sectionHeff}).

The open system dynamics of the atom-field density operator $\rho(t)$ is obtained by a numerical integration of the microscopic Markovian master equation \cite{red1}
\beq\label{me}
\frac{d\rho}{dt}=-\frac{i}{\hbar}[H(t),\rho]+\mathcal{L}_a[\rho]+\mathcal{L}_f[\rho].
\eeq
$\mathcal{L}_{a(f)}$ is the Liouvillian superoperator describing the interaction between the atom (field) with its respective thermal reservoir (see details in Appendix \ref{secME}).
The coupling between the atom (cavity) and its thermal reservoir is characterized by the decay rate $\Gamma$ (resp. $\kappa$).

%%%%%%%%%%%%%%%%%%%%%%%%%%%%%%%%%%%%%%%%%%%%%%%%%%%%%%%%%%%%%%
\subsection{Otto cycle in nonstationary cavity-QED}
\label{II.B}

\begin{figure}[!th]
\centering
\includegraphics[width=0.98\linewidth]{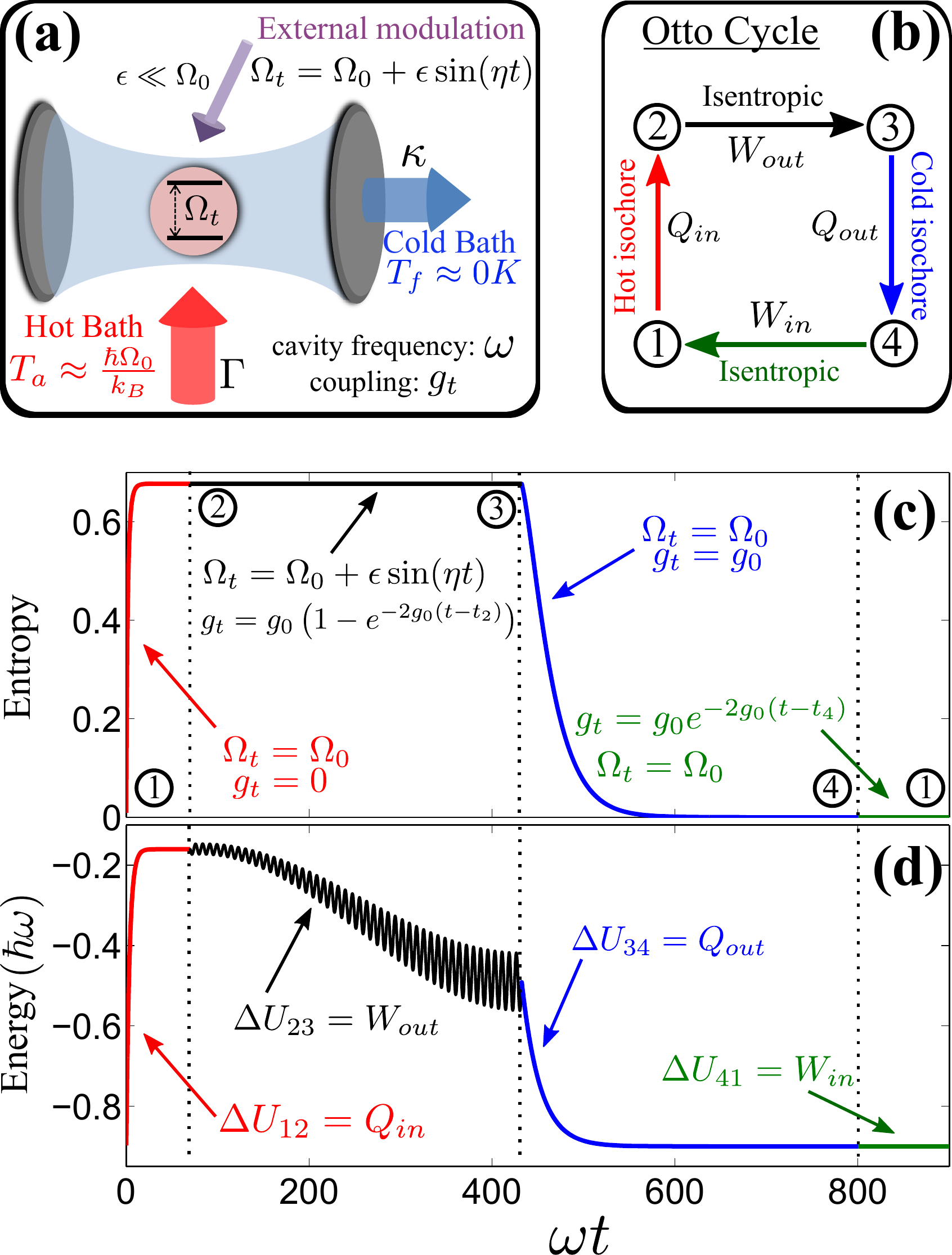}
\caption{(Color online)
{\bf Otto cycle in nonstationary cavity-QED.}
{\bf(a)} Scheme of the cavity-QED quantum thermal machine.
Working medium: two-level atom (frequency $\Omega_t$) interacts (coupling $g_t$) with a cavity (frequency $\omega$).
The atom interacts with a hot bath of temperature $T_a = 2.8 \hbar\Omega_0/k_B$.
The cavity interacts with a cold bath of temperature $T_f = 0$.
The decay rate is $\Gamma$ for the atom and  $\kappa$ for the cavity.
{\bf(b)} Four-stroke Otto cycle.
Hot isochore ($1 \rightarrow 2 $): $\Gamma=0.05\omega$, $\kappa=0$.
Isentropic work extraction ($2\rightarrow3$): $\kappa=\Gamma=0$.
Cold isochore ($3\rightarrow4$): $\Gamma=0$, $\kappa=0.05\omega$.
Isentropic reset ($4\rightarrow1$): $\kappa=\Gamma=0$.
{\bf(c)} Entropy $S(t)=-\mbox{Tr}[\rho(t)\ln\rho(t)]$ of the atom-cavity system through the Otto cycle as a function of time.
{\bf(d)} Internal energy $U(t) = \mbox{Tr}[\rho(t) H(t)]$ of the atom-cavity system through the Otto cycle as a function of time.
Heat from the hot bath, $Q_{in}$, is partly converted into work, $W_{out}$, partly wasted as heat to the cold bath, $Q_{out}$, and the cycle reset consumes $W_{in} = 0$ of work.
Parameters in (c) and (d):
$\Omega_0=1.8\omega$,
$\Delta_- \equiv \omega - \Omega_0 = - 0.8 \omega$,
$\eta = \eta_r \approx |\Delta_-|$,
$g_0=0.05\omega$,
$\epsilon=0.144\omega$.}
\label{model}
\end{figure}

%%%%%%%%%%%%%%%%%%%%%%%%%%%%%%%%%%%%%%%%%%%%%%%%%%%%%%%%%%%%%\

Now we illustrate how the nonstationary cavity-QED system, described by the JC Hamiltonian (\ref{hjc}), can be employed as the working medium of a thermal machine (see Fig. \ref{model}(a)).
A thermal machine is a device that operates cyclically between two thermal reservoirs, producing useful work with the heat extracted from hot reservoir \cite{alicki79, Kosloff2015}.
In our case, the thermal reservoirs of temperatures $T_a$ and $T_f$ are respectively coupled to the atom and to the cavity field.
We assume $T_f = 0$ (cold bath) and $T_a= 2.8 \hbar \Omega_0 /k_B$ (hot bath), where
$k_B$ is the Boltzmann constant.
We devise an Otto cycle, consisting of the four strokes described below (see Fig.\ref{model}(b)).

\

{\bf First stroke: hot isochoric.}
This process is responsible for the thermalization of the atom.
The working medium is prepared initially in the ground state,
$
\ket{g,0}
$, and one sets $\Omega_t=\Omega_0$ and $g_t = 0$.
In this stroke the atom is coupled to the hot bath, $\Gamma \neq 0$, while the cavity is kept isolated, $\kappa=0$.
The hot bath thermalizes the atom,
\beq
\ket{g,0} \rightarrow p\ket{g,0}\bra{g,0} + (1-p)\ket{e,0}\bra{e,0},
\label{hotiso}
\eeq
where $p=1/\left(1+\exp({-\hbar \Omega_0/k_BT_a})\right) \approx 0.6$.
Figs. \ref{model}(c) and (d) show that the increase of the system internal energy, $U(t) \equiv \mbox{Tr}(\rho(t) H(t))$, is accompanied by an increase of the system's entropy, $S(t) \equiv -\mbox{Tr}\left(\rho(t)\ln\rho(t)\right)$, since this stroke is a nonunitary process.
The internal energy variation, $\Delta U_{12} = U(t_2) - U(t_1)$, is exclusively due to the heat supplied by the hot bath, $Q_{in}$,
\beq
\Delta U_{12} = Q_{in} > 0.
\eeq
$t_1$ and $t_2$ are, respectively, the initial and the final times of the first stroke.
The quantum thermodynamical definitions of work and heat are discussed in Appendix \ref{secQT}.

\

{\bf Second stroke: isentropic work extraction.}
The atom-field system is isolated from both the hot and the cold baths ($\Gamma=\kappa=0$).
The atom-cavity coupling is monotonically switched on as
$g_t=g_0\left(1-\exp\left[-2g_0(t-t_2)\right]\right)$.
The atomic transition frequency also undergoes an external modulation, as described by Eq. (\ref{atf}), with the modulation frequency
\beq
\eta_r \equiv \sqrt{\Delta_-^2+4g_0^2}.
\eeq
We set $g_0=0.05\omega$ and $\Delta_- = - 0.8\omega$.
The increase of $g_t$ happens much faster than the time interval of the second stroke, $t_3 - t_2$, so the dynamics is dominated by the variation of $\Omega_t$.
Since $\Omega_0 > \omega$, the external modulation extracts energy from the atom-cavity system via the transition
\beq
\ket{e,0} \rightarrow \ket{g,1}.
\eeq
We adjust $t_3$ in order to maximize the population of the state $\ket{g,1}$.
This maximizes energy extraction, corresponding to the output work
\beq
\Delta U_{23} = W_{out} < 0.
\eeq
The value of $W_{out}$ is discussed in the following section.

\

{\bf Third stroke: cold isochoric.}
The atom-cavity system is coupled to the cold bath, $\kappa\neq0$ and $\Gamma=0$.
The system parameters remain constant over time, $\Omega_t=\Omega_0$ and $g_t=g_0$.
The cold bath ($T_f = 0$) thermalizes the coupled atom-field system to the ground state of the time-independent JC Hamiltonian,
\beq
\rho(t_4) = \ket{g,0} \bra{g,0},
\eeq
where $t_4$ is the final time of the third stroke.
During this process energy $\Delta U_{34}$ is transferred from the working medium to the cold bath in the form of heat,
\beq
\Delta U_{34} = Q_{out} < 0.
\eeq
As shown in Fig. \ref{model}(d), $|Q_{out}| < |Q_{in}|$.
This is expected, since part of the heat provided in the first stroke was converted into useful work.

\

{\bf Fourth stroke: isentropic reset.}
In the last stroke the atom-cavity system is again decoupled from the two baths,
$\Gamma=\kappa=0$.
The atomic transition frequency remains constant over time, $\Omega_t = \Omega_0$, while
the atom-cavity coupling is monotonically switched off,
\beq
g_t=g_0\exp\left[-2g_0(t-t_4)\right].
\eeq
This temporal modulation of $g_t$ is unable to perform work since the atom-cavity system remains in the ground state during this stage,
\beq
\ket{g,0} \rightarrow \ket{g,0}.
\eeq
Therefore, the variation of the internal energy is
\beq
\Delta U_{41} = W_{in} = 0,
\eeq
so the total amount of work of the Otto cycle coincides with the work extracted during the second stroke.
Fig. \ref{model}(d) evidences that
\beq
Q_{in} + W_{out} + Q_{out} + W_{in} = 0,
\eeq
as the signature of the first law of thermodynamics.

%%%%%%%%%%%%%%%%%%%%%%%%%%%%%%%%%%%%%%%%%%%%%%%%%%%%%%%%%%%%%%%%%%%
\subsection{Quantum power boost in nonstationary Jaynes-Cummings model}
\label{II.C}

\begin{figure}[!htb]
\centering
\includegraphics[width=0.98\linewidth]{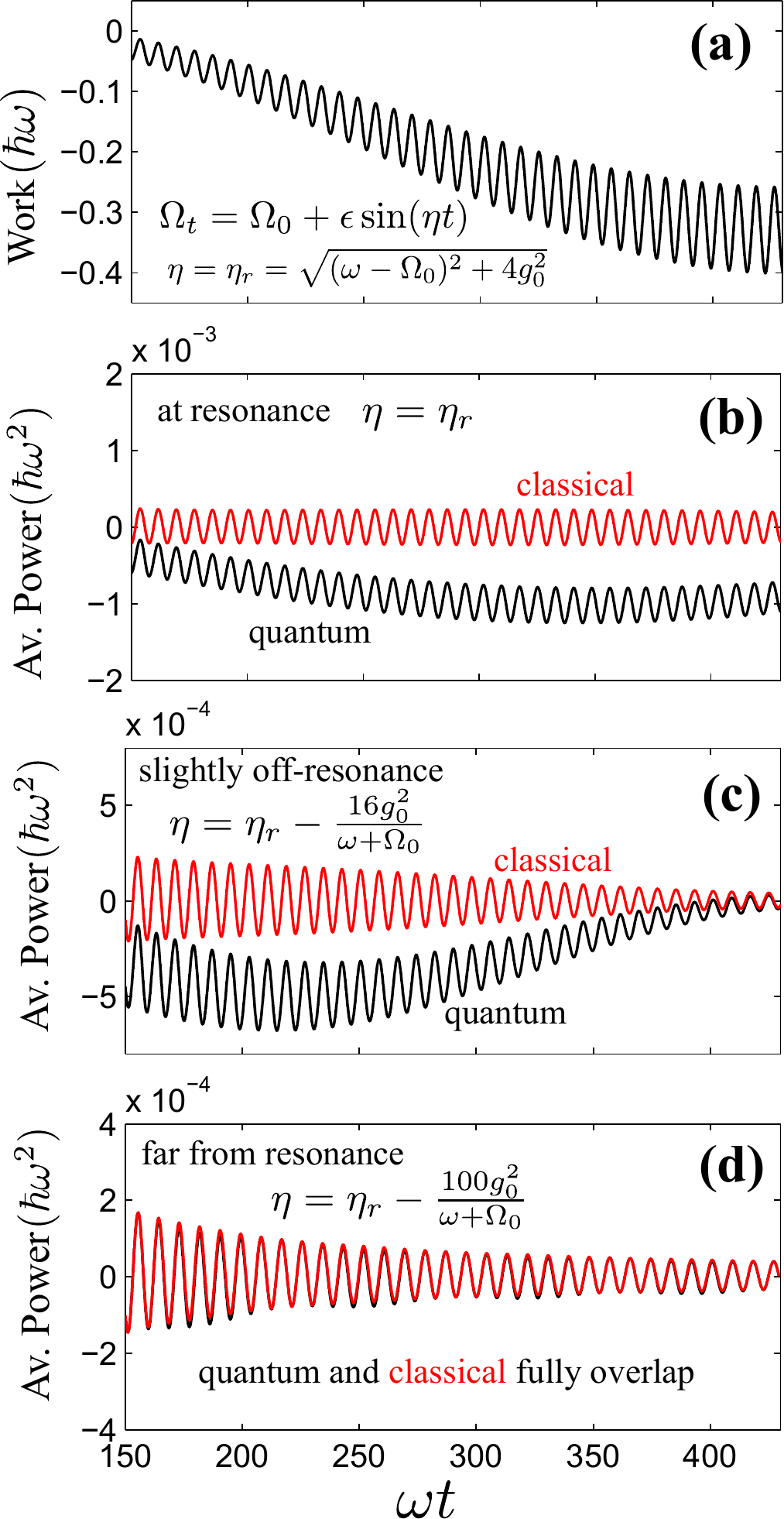}
\caption{(Color online)
{\bf Quantum power boost in the nonstationary Jaynes-Cummings model.}
{\bf(a)} Quantum work extracted during the second stroke of Sec. \ref{II.B}.
Parameters: $\Omega_0=1.8\omega$, $g_0=0.05\omega$, $\epsilon=0.144\omega$, $\eta_r\approx 0.806\omega$.
{\bf(b)} $P_{av}$ (black line) and $P^c_{av}$ (red line) at resonance, $\eta=\eta_r$.
{\bf(c)} $P_{av}$ (black line) and $P^c_{av}$ (red line) slightly off-resonance, $\eta \approx 0.792 \omega$.
{\bf(d)} $P_{av}$ (black line) and $P^c_{av}$ (red line) far from resonance, $\eta \approx 0.717 \omega$.
The quantum boost in the output power is evidenced by $P_{av} < P^c_{av}$, with $P_{av} < 0$.
As $\eta$ moves away from $\eta_r$, $P_{av}$ approaches $P^c_{av}$ due to decrease in quantum coherences.
}
\label{boostJCH}
\end{figure}

Now we seek a quantum signature in the output power from our thermal machine.
The instantaneous quantum power generated by a thermal machine is defined as
$P(t) = \mbox{Tr}\left[\rho(t)\frac{\partial H}{\partial t}\right]$.
The work extracted during the second stroke of the Otto cycle can, therefore, be obtained as
\beq
W(t)=U(t)-U(t_2)=\int_{t_2}^t \mbox{Tr}\left[\rho(\tau)\frac{\partial H}{\partial \tau}\right] \ d\tau.
\eeq
%$W(t)$ depends on the duration time $t - t_2$ of the second stroke.
Fig. \ref{boostJCH}(a) shows $W(t)$ when $\eta=\eta_r$.
As discussed above, we have adjusted the value of final time $t_3$ to ensure the maximal work extraction: that happens at
$t_3 - t_2=\pi/(2\lambda) \approx \pi |\Delta_-| / (g_0\epsilon)$ ($\lambda$ is the transition rate between the involved states, see Appendix \ref{sectionHeff}).
%$t_3 \sim \lambda^{-1}$ (see Appendix A for details).
The amount of work withdrawn from the system is then
$W(t_3) \approx (1-p) \ \hbar \Delta_- \approx -0.3 \hbar \omega$,
since $1-p \approx 0.4$, as given below Eq. (\ref{hotiso}), and $\Delta_- = - 0.8\omega$.
The fast oscillations observed in Fig. \ref{boostJCH}(a) have frequency of the order of the modulation frequency, $\eta_r$.

The average quantum power \cite{Seifert2017},
\beq\label{power}
P_{av} \equiv \frac{1}{t-t_2} \int_{t_2}^t \left( \mbox{Tr}\left[\rho(\tau)\frac{\partial H}{\partial \tau}\right] \right) d\tau,
\eeq
will be compared to the average classical power \cite{Seifert2017},
%Therefore, the average classical power generated per cycle associated with the corresponding classical thermal machine is given by \cite{Seifert2017}:
\beq\label{cpower}
P^c_{av} \equiv \frac{1}{t-t_2}\int_{t_2}^t \left(\sum_n \rho_{nn}(\tau) \frac{\partial E_n}{\partial \tau}\right)d\tau,
\eeq
where $\rho_{nn}(\tau) = \bra{E_n(\tau)} \rho(\tau) \ket{E_n(\tau)}$ and $H(\tau)  \ket{E_n(\tau)} = E_n(\tau) \ket{E_n(\tau)}$,
for the reason explained in the following.
According to Refs.\cite{Kosloff2015, Seifert2017}, the presence of quantum coherence in a thermal machine can lead to increased output power with respect to the power from the corresponding classical thermal machine.
Gain in the output power can happen because, in the quantum case, two physical mechanisms are responsible for the work extraction.
The first one is associated with the variation of the energy levels of the working medium, whereas the second mechanism depends on the quantum coherence between the instantaneous energy eigenstates.
In the classical case only the first physical mechanism is responsible for the work extraction, explaining why the signature of the \emph{quantum boost} in the output power appears as
\beq
P_{av} - P^c_{av} < 0.
\eeq
The more negative the difference, the larger is the quantum boost in the output power.
%Eq.(\ref{cpower}) can be obtained from Eq.(\ref{power}) by neglecting the effects of the quantum coherence of the atom-field system, that is, by using only the diagonal terms of the density operator $\rho(t)$, in the basis that instantaneously diagonalizes Hamiltonian (\ref{hjc}).
%%%%%%%%%%%%%

Figs. \ref{boostJCH}(b)-(d) reveal the physical meaning of $P_{av} - P^c_{av}$ in the context of our thermal machine.
The main message here is: the closer $\eta$ is to the resonance $\eta_r$, the more energy can be extracted by means of a quantum coherent process, surpassing classically available mechanisms due to energy levels variation.
In Fig. \ref{boostJCH}(b), we plot $P_{av}$ (black line) and $P^c_{av}$ (red line) for the second stroke of our Otto cycle, where $\eta=\eta_r \approx 0.806\omega$.
$P^c_{av}$ oscillates around zero at frequency $\sim \eta_r$, as a consequence of the variation of energy levels due to $\Omega_t$, so the external modulation both supplies energy to the system and draws energy from the system, with no net energy variation.
On the other hand, despite the same fast oscillations due to the variation of the energy levels, $P_{av}$  assumes strictly negative values.
Therefore, the power generated by our quantum thermal machine outperforms the power generated by the corresponding classical thermal machine, $\left|P_{av}\right| > \left|P^c_{av}\right|$.
In Figs. \ref{boostJCH}(c) and (d), we plot $P_{av}$ (black line) and $P^c_{av}$ (red line) for two different modulation frequencies:
$\eta=\eta_r-16g_0^2/(\omega+\Omega_0) \approx 0.792 \omega$ (panel c) and
$\eta=\eta_r-100g_0^2/(\omega+\Omega_0) \approx 0.717 \omega$ (panel d).
We find that $|\eta - \eta_r|\gg \lambda$  implies $P_{av} \approx P^c_{av}$, since the external modulation goes off-resonance and is no longer able to induce transitions between the energy eigenstates. Hence the only physical mechanism responsible for finite power output is the variation of the energy levels, with no quantum coherences being created in the energy basis.
%

%%%%%%%%%%%%%%%%%%%%%%%%%%%%%%%%%%%%%%%%%%%%%%%%%%%%%%%%%%%%%%
\subsection{Amplification of the driving field}
\label{II.D}

We now answer the question: what happens to the extracted work?
Conservation of energy implies amplification of the driving field, responsible for the variation of $\Omega_t$.
The mechanism behind the field amplification is the stimulated transition between two eigenstates  (dressed-states) of the coupled atom-cavity system, from the populated upper level $\ket{1,+}$ to the unpopulated lower level $\ket{1,-}$ (see Appendix \ref{sectionHeff}).
The frequency of the driving field is tuned to resonance with this transition,
\beq
E_{1,+} - E_{1,-} = \hbar \eta_r.
\eeq
%An effective Hamiltonian, Eq.(\ref{eqHeff}), describes the coupled-system dynamics, and is equivalent to the Hamiltonian of a two-level atom driven by an classical resonant field, with classical Rabi frequency $\lambda = ??$.
The stimulated emission of a photon amplifies the driving field by the amount
\beq
|W_{out}| \approx (p_{1,+} - p_{1,-}) (E_{1,+} - E_{1,-}),
\eeq
where $p_{1,\pm}$ is the population of the state $\ket{1,\pm}$ at the beginning of the work-extraction stroke.
In the previous sections, we have explored the case where
$\ket{1,+} \approx \ket{e,0}$,
$\ket{1,-} \approx \ket{g,1}$,
$E_{1,+} - E_{1,-} \approx \hbar |\Delta_-| = 0.8\hbar \omega$,
$p_{1,+} \approx 0.4$ and
$p_{1,-} = 0$,
yielding
$|W_{out}| \approx 0.4\times 0.8\hbar\omega \approx 0.3 \hbar \omega$, as shown in Fig. \ref{boostJCH}(a).

Field amplification due to stimulated emission at the single-photon level has been investigated both in the steady-state  \cite{sandoghdar, tsai2010} and in the transient \cite{tsai2011, dv2012A} regimes.
Refs.\cite{sandoghdar, tsai2010,tsai2011, dv2012A} show that stimulated emission can be monitored by an increase in the transmitted power $T = \langle E_T^\dagger E_T\rangle$.
In the optical regime \cite{sandoghdar,dv2012A}, $E_T$ represents the propagating electric field, whereas in the microwave regime \cite{tsai2010,tsai2011}, it represents an electric current in the circuit.
The increase of the transmitted power is due to a constructive interference,
$E_T = E_{f} + E_{\tilde{\sigma}}$,
between the free field, $E_f$, and the field emitted by the two-level system, $E_{\tilde{\sigma}} \propto \tilde{\sigma}$, as obtained from the Maxwell-Bloch theory.
Here, the ladder operator for the effective two-level system is ${\tilde{\sigma}} = \ket{1,-} \bra{1,+}$.
This explains how the extracted work amplifies the driving field by the stimulated emission of a photon.

%%%

%%%%%%%%%%%%%%%%%%%%%%%%%%%%%%%%%%%%%%%%%%%%%%%%%%%%%%%%%%%%%%
\section{Quantum power boost in the nonstationary Rabi model}
\label{III}

\begin{figure}[!htb]
\centering
\includegraphics[width=1.0\linewidth]{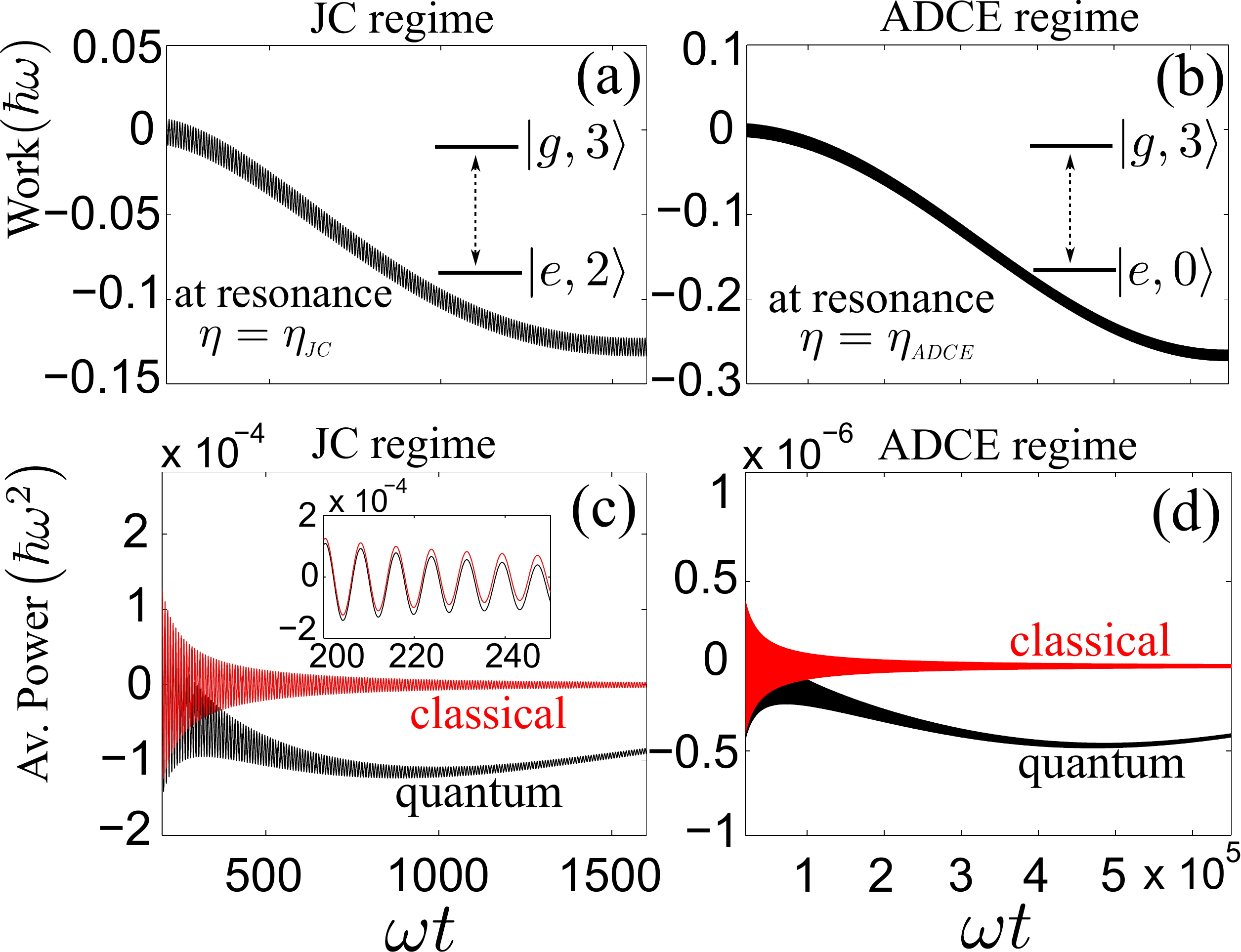}
\caption{(color online)
{\bf Quantum power boost in the nonstationary Rabi model.}
{\bf (a)} $W_{\mbox{\it \tiny{JC}}}(t)$ at resonance, $\eta_{\mbox{\it \tiny{JC}}} = 0.8021\omega$.
{\bf (b)} $W_{\mbox{\it \tiny{ADCE}}}(t)$ at resonance, $\eta_{\mbox{\it \tiny{ADCE}}} = 2.8041\omega$.
%$|W_{\mbox{\it \tiny{ADCE}}}(t_3)| > |W_{\mbox{\it \tiny{JC}}}|$, although the time required for work extraction in the JC regime is significantly lower.
{\bf (c)} $P_{av}$ (black line) and $P^c_{av}$ (red line) for the JC regime, at resonance $\eta_{\mbox{\it \tiny{JC}}}$.
{\bf (d)} $P_{av}$ (black line) and $P^c_{av}$ (red line) for the ADCE regime, at resonance $\eta_{\mbox{\it \tiny{ADCE}}}$.
$P_{av} < P^c_{av}$ and $P_{av} < 0$
are attained in both regimes of the nonstationary Rabi model (JC and ADCE), evidencing the quantum boosts in the output powers.
{\it Inset:} fast oscillations of $P_{av}$ and $P^c_{av}$ have period $\sim \eta^{-1}$.
%The behavior of the classical power is close to the quantum power in the initial time interval.
Parameters used in all plots: $\Omega_0=0.2\omega$, $g_0=0.05\omega$, $\epsilon=0.016\omega$.
}
\label{boostRabi}
\end{figure}

In this section we show that the quantum power boost discussed in Sec. \ref{II.C} for the JC model exhibits a similar behavior in the nonstationary Rabi model.
The Rabi Hamiltonian reads \cite{jpcs,jpa,red1,AD2017,RabiCircuitQED}
\beq\label{hr}
H_R(t)=\omega a^\dagger a +\frac{\Omega_t}{2}\sigma_z + g_t(\sigma_+ +\sigma_-)(a+a^\dagger)
\eeq
and it differs from the Jaynes-Cummings Hamiltonian by the presence of the counter-rotating terms (CRT) $a\sigma_-$ and $a^\dagger\sigma_+$.
The existence of the counter-rotating terms implies that the number of excitations of the system is not a conserved quantity.
Therefore the system can undergo transitions between states $\ket{g,n}$ and $\ket{e,n+1}$, as well as the transition between the states $\ket{g,n}$ and $\ket{e,n-1}$ allowed by the JC Hamiltonian.

For $g_0\ll\omega$ the effects of the counter-rotating terms usually appear when a \lq high\rq\ modulation frequency $\eta\sim 2\omega$ becomes resonant with some transition between the dressed-states with different numbers of excitations. For the transition $|g,n+1 \rangle\leftrightarrow|e,n\rangle$ studied in Sec. \ref{II} the CRT simply shift the resonance frequency to \cite{palermo,red1}
\beq
\eta_{\mbox{\it \tiny{JC}}}=\sqrt{[\Delta_--2\delta_+ (n+1)]^2+4g_0^2 (n+1)},
\label{dave}
\eeq
where $\delta_+=g_0^2/(\omega+\Omega_0)$ is the standard Bloch-Siegert shift.
In this case, that we call JC regime, only transitions between the states $\ket{g,n+1}$ and $\ket{e,n}$ take place, and the dynamics is (approximately) described by the same effective Hamiltonian as discussed in Sec. \ref{II}.

We have shown in our previous work \cite{AD2017} that additional transitions induced by the counter-rotating terms can be used to increase the amount of work extracted from the atom-cavity system, by exploring the dynamic variation of the total number of excitations.
The ADCE regime, for instance, promotes coherent annihilation of two system excitations.
By adjusting the modulation frequency as \cite{fn}
\beq
\eta_{\mbox{\it \tiny{ADCE}}} \approx 3 \omega-\Omega_0
\label{mustaine}
\eeq
one can engineer an effective dynamics that couples the (approximate) states \cite{igor,diego}
\beq
\ket{g,n} \ \ \mbox{and} \ \ \ket{e,n-3}, \ \ \mbox{for} \ \ n\geq3.
\eeq
To be able to reduce the number of system excitations it is required the evident condition \cite{AD2017}
\beq
p(\ket{g,n}) > p(\ket{e,n-3}),
\label{invertedpop}
\eeq
where $p(\ket{.})$ is the initial population of the state $\ket{.}$.

In the context of quantum thermal machines, the requirement (\ref{invertedpop}) can be achieved by an isochoric process similar to that employed in the first stroke of our Otto cycle (Sec. \ref{II.B}).
The difference now is that the hot reservoir is coupled to the cavity ($\kappa > 0$ and $\Gamma = 0$), thermalizing the cavity field at a finite temperature, $T_f > 0$.
The atom remains in its ground state, since $g_{t} = 0$ for $t_1 < t < t_2$.
The atom-cavity state at the beginning of the work-extraction stroke is then given by
\beq
\rho(t_2) = \ket{g} \bra{g} \otimes \sum_n p_n \ket{n} \bra{n},
\eeq
where
$p_n = \bar{n}^n/(\bar{n}+1)^{n+1}$ and $\bar{n}$ is the average photon number.
The next step is to implement the work extraction process described in the second stroke of the Otto cycle, using state $\rho(t_2)$ as the initial state.

Figs. \ref{boostRabi}(a) and (b), respectively, show work extracted in the JC regime, $W_{\mbox{\it \tiny{JC}}}(t)$, and in the ADCE regime,  $W_{\mbox{\it \tiny{ADCE}}}(t)$, of the nonstationary Rabi model.
In both regimes we set $\bar{n}=1.8$ and $\Omega_0 = 0.2\omega$ (so now $\Delta_- = +0.8\omega$).
We also set $\eta_{\mbox{\it \tiny{JC}}}=1.0026\Delta_-$ to select the transition
\beq
\ket{g,3} \rightarrow \ket{e,2},
\eeq
and
$\eta_{\mbox{\it \tiny{ADCE}}}=1.0015(3\omega-\Omega_0)$,
which induces transition
\beq
\ket{g,3} \rightarrow \ket{e,0}.
\eeq
Qualitatively, $W_{\mbox{\it \tiny{JC}}}(t)$ and $W_{\mbox{\it \tiny{ADCE}}}(t)$ are equivalent.
From a quantitative point of view, however, the results show that maximal work extraction in the ADCE regime can be twice as large as in the JC regime, as allowed by the reduction in the total number of excitations in the atom-cavity system.
Maximal work extraction in the ADCE regime occurs when the population of the state $\ket{g,3}$ attains its minimum value, due to the transfer of population to the state $\ket{e,0}$.
The time required to reach this stage in the ADCE regime is significantly larger than in the JC regime \cite{AD2017}, hence the average quantum power extracted in the JC regime (Fig. \ref{boostRabi}(c), black line) is much higher than in the ADCE regime (Fig. \ref{boostRabi}(d), black line).

Figs. \ref{boostRabi}(c) and (d) illustrate quantum boosts in the output powers for the JC and the ADCE regimes of the nonstationary Rabi model.
In both regimes $P^c_{av}$ oscillates around zero due to the modulation of energy levels (red lines in panels c and d).
These oscillations are characterized by a very short period ($\sim\eta^{-1}$), as illustrated in the inset of Fig. \ref{boostRabi}(c).
Moreover, in both regimes $P_{av}$ becomes strictly negative (see black lines in panels c and d).
Finally, we have confirmed the role of quantum coherence on the quantum power boost by slightly changing the modulation frequency $\eta$, thereby weakening the transitions between the required dressed-states (data not shown).
Similarly to the results from Figs. \ref{boostJCH}(b)-(d), we observed that in both the JC and ADCE regimes $P_{av}$ approaches $P^c_{av}$ as the modulation frequency moves away from resonance.
%The contribution of quantum coherence can also be observed in the {\it inset} of the Fig.3(c).
%The behavior of the classical power is close to the quantum power in the initial time interval.
%This result can be understood in view of the fact that the time scale associated with the variations of energy levels ($\sim\eta^{-1}$) is much smaller than the time scale associated with the transition $\ket{g,3}\longleftrightarrow\ket{e,2}$ ($\sim\Delta_-/g_0\epsilon_\Omega)$, which is responsible for creating quantum coherence between energy eigenstates.

%%%%%%%%%%%%%%%%%%%%%%%%%%%%%%%%%%%%%%%%%%%%%%%%%%%%%%%%%%%%%%
\section{Conclusions}
\label{Conclusions}

We have shown how quantum coherences in nonstationary cavity-QED regimes can boost the output power of a quantum heat engine.
We have evidenced the quantum boost by comparing the average quantum power, $P_{av}$, with the average classical power, $P^c_{av}$, both as defined in \cite{Seifert2017}.
$P_{av}$ captures two mechanisms: quantum coherent transitions between energy levels and the time variation of the energy levels themselves.
$P^c_{av}$ only captures the time variation of the energy levels, representing the average power of the equivalent classical thermal machine \cite{Kosloff2015}, so the quantum boost in the extracted power reveals itself as $P_{av} < P^c_{av}$, with $P_{av} < 0$.
We have shown that such quantum boost can be achieved by adjusting the modulation frequency $\eta$ to specific resonant frequencies that induce transition between specific pairs of the system dressed-states. 
For the nonstationary Jaynes-Cummings model, we devised a four-stroke Otto cycle and explained the amplification of the driving field by means of stimulated emission at the single-photon level. 
Finally, for the nonstationary Rabi model, we demonstrated the quantum power boost in both the JC and ADCE regimes.

%

%%%%%%%%%%%%%%%%%%%%%%%%%%%%%%%%%%%%%%%%%%%%%%%%%%%%%%%%%%%%%
%%%%%%%%%%%%%%%%%%%%%%%%%%%%%%%%%%%%%%%%%%%%%%%%%%%%%%%%%%%%%
\appendix

\section{Effective Hamiltonian}
\label{sectionHeff}
Under external driving (\ref{atf}) with modulation frequency (\ref{nstres}), the effective Hamiltonian for the nonstationary Jaynes-Cummings model (\ref{hjc}) is given by \cite{igor, AD2017,palermo}
\beq
H_{\mathrm{eff}}/\hbar\simeq i\lambda\ket{n+1,-}\bra{n+1,+}+h.c.\ ,
\label{eqHeff}
\eeq
where $n\ge 0$ and
\beq
\lambda = \frac{g_0 \epsilon}{2|\Delta_-|} \sqrt{n+1} .
\eeq
For $m>0$
\begin{eqnarray}
\ket{m,+}&=&\sin\theta_m\ket{g,m}+\cos\theta_m\ket{e,m-1}\\
\ket{m,-}&=&\cos\theta_m\ket{g,m}-\sin\theta_m\ket{e,m-1}
\end{eqnarray}
are the $m$-excitations eigenstates of the Hamiltonian (\ref{hjc}) with $\theta_{m}=\arctan\left[(\Delta_-+\beta_m)/2g_0\sqrt{m}\right]$, $\Omega_t=\Omega_0$ and $g_t=g_0$.
The respective eigenenergies are
\beq
E_{m,\pm}/\hbar= \omega(m-1/2)\pm\frac{\beta_m}{2} ,
\eeq
with $\beta_m=\sqrt{\Delta_-^2+4g_0^2m}$.
In the dispersive regime $\ket{m,s}\approx\ket{g,m}+(g_0\sqrt{m}/\Delta_-)\ket{e,m-1}$ and $\ket{m,-s}\approx\ket{e,m-1}-(g_0\sqrt{m}/\Delta_-)\ket{g,m}$, where $s\equiv sgn(\Delta_-)$.
%{\bf Falar brevemente que este nao e um toy model - mencionar a secao III}.

\

\section{Master Equation}
\label{secME}
In this paper we use the microscopic Markovian master equation developed in Ref. \cite{red1}, which takes into account the influence of the atom-field interaction on the description of the dissipative effects. The Liouvillian superoperators read
\begin{eqnarray}
\mathcal{L}_f[\rho] &=& \sum_{j,k>j} \  \kappa^{jk} \ (n_f(\Delta_{kj},T_f) + 1) \ \mathcal{D}[\ket{j} \bra{k} ]\rho \nn\\
&+& \sum_{j,k>j} \ \kappa^{jk} \ n_f(\Delta_{kj},T_f) \ \mathcal{D}[\ket{k} \bra{j} ]\rho
\end{eqnarray}
\begin{eqnarray}
\mathcal{L}_a[\rho] &=& \sum_{j,k>j} \ \Gamma^{jk} \ (n_a(\Delta_{kj},T_a)+1) \ \mathcal{D}[\ket{j} \bra{k} ]\rho  \nn\\
&+& \sum_{j,k>j} \ \Gamma^{jk} \ n_a(\Delta_{kj},T_a) \ \mathcal{D}[\ket{k} \bra{j} ]\rho,
\end{eqnarray}
where
$\mathcal{D}[O]\rho =  (2O\rho O^\dagger - \rho O^\dagger O - O^\dagger O\rho)/2$ is the Lindbladian superoperator and the shorthand notation $\ket{k}$ stands for the eigenstates of the JC Hamiltonian, where the index $k$ increases with the eigenenergy $E_{k}$.
Other parameters are defined
as  $\kappa^{jk}=\kappa (\Delta _{kj})|a^{jk}|^{2}$ and $\Gamma^{jk}=\Gamma (\Delta _{kj})|\sigma _{x}^{jk}|^{2}$, where $\kappa (\varpi )$ and $\Gamma (\varpi )$ are the dissipation
rates proportional to noise spectral densities (at frequency $\varpi $) for the cavity and the atom, respectively. Moreover, $\Delta _{kj}\equiv E _{k}-E_{j} $, $a^{jk}\equiv\langle j|(\hat{a}+\hat{a}^{\dagger })|k\rangle $ and $\sigma_{x}^{jk}\equiv\langle j|(\hat{\sigma}_{+}+\hat{\sigma}_{-})|k\rangle $. We make the simplest assumption that the dissipation rates are
zero for $\varpi <0$ and take on constant values $\kappa $ and $\Gamma$ for $\varpi \geq 0$. Finally, $n_{a(f)}$ denotes the mean number of excitations associated with the thermal reservoir coupled to the atom (cavity field) at temperatures $T_{a(f)}$ and energy difference $\Delta_{kj}$.
It is worth mentioning that in cases where the system of interest interacts with reservoirs at different temperatures the microscopic approach is applicable, whereas the phenomenological approach may lead to violation of the second law of thermodynamics \cite{kosloff14}.

\

\section{Quantum Thermodynamics}
\label{secQT}
In quantum thermodynamics of open quantum systems, the internal energy $U$ of the system of interest is associated with its average energy, $U(t)=\mbox{Tr}[\rho(t) H(t)]$, where $\rho(t)$ is the density operator and $H(t)$ is the system Hamiltonian.
A thermodynamic process is represented by the temporal evolution of the system from time $t_i$ up to $t_f$.
The quantum version of the first law of thermodynamics can be formulated as follows \cite{alicki79,Kosloff2015,AD2017}
\beq
\Delta U=W+Q,
\eeq
where
\beq
W=\int_{t_i}^{t_f}\mbox{Tr}[\rho(t) \partial_tH(t)]\ dt
\eeq
is the quantum version of the work performed by an external agent during the thermodynamic process.
For a time-independent Hamiltonian, $W=0$.
The quantum version of the heat is
\beq
Q=\int_{t_i}^{t_f}\mbox{Tr}[H(t)\partial_t\rho(t) ] \ dt.
\eeq
The quantum heat depends on the temporal variation of the system density operator, which is given by a master equation as in Eq. (\ref{me}).

In our case, using the Hamiltonians (\ref{hjc}) and (\ref{hr}), for $g_t=g_0$ the quantum work takes the form
\beq
W=\frac{1}{2}\int_{t_i}^{t_f}\partial_t\Omega_t\left\langle \sigma_z(t)\right\rangle dt
\eeq
(in the numeric simulations the time dependence of $g_t$ is also taken into account, although its contribution is negligible).
Using Eq. (\ref{me}) and that $\mbox{Tr}\left\{ [H(t),\rho(t)] H(t) \right\} = 0$,
the quantum heat can be rewritten as $Q=Q_a+Q_f$, where
\beq
Q_{a(f)}=\int_{t_i}^{t_f}\mbox{Tr}[\mathcal{L}_{a(f)}\rho H(t)] \ dt
\eeq
is the heat transferred into the system by the thermal reservoir coupled to the atom (cavity field).

%%%%%%%%%%%%%%%%%%%%%%%%%%%%%%%%%%%%%%%%%%%%%%%%%%
%
\begin{acknowledgments}
A. V. D. acknowledges partial support from the Brazilian agency Conselho Nacional de Desenvolvimento Cient\'ifico e Tecnol\'ogico (CNPq).
D. V. acknowledges support from the Brazilian agency Coordena\c c\~ao de Aperfei\c coamento de Pessoal de N\'ivel Superior (CAPES) through grant No. 88881.120135/2016-01.
D. V. and T. W. acknowledge support from Instituto Nacional de Ci\^encia e Tecnologia - Informa\c c\~ao Qu\^antica (INCT-IQ), Brazil.
\end{acknowledgments}

%%%%%%%%%%%%%%%%%%%%%%%%%%%%%%%%%%%%%%%%%%%%%%%%%%
%
%


\begin{thebibliography}{99}

%\bibitem{review} Review;

\bibitem{Kosloff2015} R. Uzdin, A. Levy and R. Kosloff,
% {\it Equivalence of Quantum Heat Machines, and Quantum-Thermodynamic Signatures},
Phys. Rev. X \textbf{5}, 031044 (2015).

\bibitem{Scully2011} M. O. Scully, K. R. Chapin, K. E. Dorfman, M. B. Kim, and A. Svidzinsky,
% {\it Quantum heat engine power can be increased by noise-induced coherence}
Proc. Nat. Acad. Sci. \textbf{108}, 15097 (2011).

\bibitem{Kurizki2015} D. Gelbwaser-Klimovsky, W. Niedenzu, P. Brumer, and G. Kurizki,
% {\it Power enhancement of heat engines via correlated thermalization in a three-level working fluid},
Sci. Rep. \textbf{5}, 14413 (2015).

\bibitem{Walmsley2017} J. Klatzow, C. Weinzetl, P. M. Ledingham, J. N. Becker, D. J. Saunders, J. Nunn, I. A. Walmsley, R. Uzdin, and E. Poem,
% {\it Experimental demonstration of quantum effects in the operation of microscopic heat engines},
arXiv:1710.08716v2.
%An ensemble of nitrogen-vacancy centers in diamond has been used to show measured output powers that beat the stochastic bound by four standard deviations and that decrease below the bound as coherence is reduced.


\bibitem{Seifert2017} K. Brandner, M. Bauer and U. Seifert,
% {\it Universal Coherence-Induced Power Losses of Quantum Heat Engines in Linear Response},
Phys. Rev. Lett. \textbf{119}, 170602 (2017).

\bibitem{Pekola2016} B. Karimi and J. P. Pekola,
% {\it Otto refrigerator based on a superconducting qubit-Classical and quantum performance},
Phys. Rev. B \textbf{94}, 184503 (2016).

\bibitem{q1} C. Sayrin, I. Dotsenko, X. Zhou, B. Peaudecerf, Th. Rybarczyk, S. Gleyzes, P. Rouchon, M. Mirrahimi, H. Amini, M. Brune, J.-M. Raimond, and S. Haroche, Nature \textbf{477}, 73 (2011).

\bibitem{q2} G. G\"{u}nter, A. A. Anappara, J. Hees, A. Sell, G. Biasiol, L. Sorba, S. De Liberato, C. Ciuti, A. Tredicucci, A. Leitenstorfer, and R. Huber, Nature \textbf{479}, 376 (2011).

\bibitem{q3} A. Stockklauser, P. Scarlino, J. V. Koski, S. Gasparinetti, C. K. Andersen, C. Reichl, W. Wegscheider, T. Ihn, K. Ensslin, and A. Wallraff, Phys. Rev. X \textbf{7}, 011030  (2017).

\bibitem{q4} P. E. Barclay, C. Santori, K.-M. Fu, R. G. Beausoleil, and O. Painter, Opt. Express \textbf{17}, 8081 (2009).

\bibitem{q5} F. Brennecke, T. Donner, S. Ritter, T. Bourdel, M. K\"{o}hl, and T Esslinger, Nature \textbf{450}, 268 (2007).

\bibitem{q6} Y. Colombe, T. Steinmetz, G. Dubois, F. Linke, D. Hunger, and J. Reichel, Nature \textbf{450}, 272 (2007).

\bibitem{you} J. Q. You and F. Nori, Nature \textbf{474}, 589 (2011).

\bibitem{nori} P. D. Nation, J. R. Johansson, M. P. Blencowe, and F. Nori, Rev. Mod. Phys. \textbf{84}, 1 (2012).

\bibitem{science} M. H. Devoret and R. J. Schoelkopf, Science \textbf{339}, 1169
(2013).

\bibitem{rev1} S. Schmidt and J. Koch, Ann. Phys. (Berlin) \textbf{525},
395 (2013).

\bibitem{nori2017} X. Gu, A. F. Kockum, A. Miranowicz,Yu-xi Liu, and F.
Nori, Phys. Rep. \textbf{718-719}, 1 (2017).

\bibitem{jpcs} A. V. Dodonov, J. Phys.: Conf. Ser. \textbf{161}, 012029 (2009).
%Photon creation from vacuum and interactions engineering in nonstationary circuit QED

\bibitem{jpa} A. V. Dodonov, J. Phys. A: Math. Theor. \textbf{47}, 285303 (2014).
%Analytical description of nonstationary circuit QED in the dressed-states basis

\bibitem{majer} J. Majer, J. M. Chow, J. M. Gambetta, J. Koch, B. R.
Johnson, J. A. Schreier, L. Frunzio, D. I. Schuster, A. A. Houck, A.
Wallraff, A. Blais, M. H. Devoret, S. M. Girvin, and R. J. Schoelkopf,
%Coupling superconducting qubits via a cavity bus, 
Nature \textbf{449}, 443 (2007).

\bibitem{ge} M. Hofheinz, H. Wang, M. Ansmann, R. C. Bialczak, E. Lucero, M.
Neeley, A. D. O'Connell, D. Sank, J. Wenner, J. M. Martinis, and A. N.
Cleland, 
%Synthesizing arbitrary quantum states in a superconducting resonator, 
Nature \textbf{459}, 546 (2009).

\bibitem{ger} L. DiCarlo, J. M. Chow, J. M. Gambetta, L. S. Bishop, B. R.
Johnson, D. I. Schuster, J. Majer, A. Blais, L. Frunzio, S. M. Girvin, and
R. J. Schoelkopf, 
%Demonstration of two-qubit algorithms with a superconducting quantum processor, 
Nature \textbf{460}, 240 (2009).

\bibitem{ger1} J. Li, M. P. Silveri, K. S. Kumar, J. -M. Pirkkalainen, A.
Veps\"{a}l\"{a}inen, W. C. Chien, J. Tuorila, M. A. Sillanp\"{a}\"{a}, P. J.
Hakonen, E. V. Thuneberg, and G. S. Paraoanu, 
%Motional averaging in a superconducting qubit, 
Nat. Commun. \textbf{4}, 1420 (2013).

\bibitem{v1} S. J. Srinivasan, A. J. Hoffman, J. M. Gambetta, and A. A.
Houck, 
Phys. Rev. Lett. \textbf{106}, 083601 (2011).

\bibitem{v2} Y. Chen, C. Neill, P. Roushan, N. Leung, M. Fang, R. Barends,
J. Kelly, B. Campbell, Z. Chen, B. Chiaro, A. Dunsworth, E. Jeffrey, A.
Megrant, J. Y. Mutus, P. J. J. O'Malley,
C. M. Quintana, D. Sank, A. Vainsencher, J. Wenner, T.
C. White, M. R. Geller, A. N. Cleland, and J. M. Martinis, 
Phys. Rev. Lett. \textbf{113}, 220502 (2014).

\bibitem{v3} S. Zeytino\u{g}lu, M. Pechal, S. Berger, A. A. Abdumalikov,
Jr., A. Wallraff, and S. Filipp, Phys. Rev. A \textbf{91}, 043846 (2015).

\bibitem{nori-n} C. M. Wilson, G. Johansson, A. Pourkabirian, M. Simoen, J.
R. Johansson, T. Duty, F. Nori, and P. Delsing, Nature \textbf{479}, 376 (2011).

\bibitem{meta} P. L\"{a}hteenm\"{a}ki, G. S. Paraoanu, J. Hassel, and P. J.
Hakonen, Proc. Nat. Acad. Sci. \textbf{110}, 4234 (2013).

\bibitem{blais-exp} J. D. Strand, M. Ware, F. Beaudoin, T. A. Ohki, B. R.
Johnson, A. Blais, and B. L. T. Plourde, Phys. Rev. B \textbf{87},
220505(R) (2013).

\bibitem{simmonds} M. S. Allman, J. D. Whittaker, M.
Castellanos-Beltran, K. Cicak, F. da Silva, M. P. DeFeo, F.
Lecocq, A. Sirois, J. D. Teufel, J. Aumentado, and R. W.
Simmonds, 
Phys. Rev. Lett. \textbf{112}, 123601 (2014).

\bibitem{schuster} Y. Lu, S. Chakram, N. Leung, N. Earnest, R. K.
Naik, Z. Huang, P. Groszkowski, E. Kapit, J. Koch, and D. I. Schuster, Phys. Rev. Lett.
\textbf{119}, 150502 (2017).

\bibitem{liberato} S. De Liberato, D. Gerace, I. Carusotto, and C. Ciuti, Phys. Rev. A \textbf{80}, 053810 (2009).
%Extracavity quantum vacuum radiation from a single qubit

\bibitem{fujii} T. Fujii, S. Matsuo, N. Hatakenaka, S. Kurihara, and A. Zeilinger, Phys. Rev. B \textbf{84}, 174521 (2011).

\bibitem{roberto} A. V. Dodonov, R. Lo Nardo, R. Migliore, A. Messina, and
V. V. Dodonov, J. Phys. B: At. Mol. Opt. Phys. \textbf{44},
225502 (2011).

\bibitem{entangles} S. Felicetti, M. Sanz, L. Lamata, G. Romero, G.
Johansson, P. Delsing, and E. Solano, Phys. Rev. Lett. \textbf{113}, 093602 (2014).

\bibitem{igor} I. M. de Sousa and A. V. Dodonov, J. Phys. A: Math. Theor. \textbf{48}, 245302 (2015).
%Microscopic toy model for the cavity dynamical Casimir effect

\bibitem{diego} D. S. Veloso and A. V. Dodonov, J. Phys. B: At. Mol. Opt. Phys. \textbf{48}, 165503 (2015).
%Prospects for observing dynamical and anti- dynamical Casimir effects in circuit QED due to fast modulation of qubit parameters

\bibitem{lucas} L. C. Monteiro and A. V. Dodonov, Phys. Lett. A \textbf{380}, 1542 (2016).
%Anti-dynamical Casimir effect with an ensemble of qubits

\bibitem{juan} A. V. Dodonov, J. J. D\'{\i}az-Guevara, A. Napoli, and B.
Militello, Phys. Rev. A \textbf{96}, 032509 (2017).

\bibitem{AD2017} A. V. Dodonov, D. Valente and T. Werlang,
%{\it Antidynamical Casimir effect as a resource for work extraction},
Phys. Rev. A \textbf{96}, 012501 (2017).

\bibitem{red2} F. Beaudoin, M. P. da Silva, Z. Dutton, and A. Blais, Phys. Rev. A \textbf{86}, 022305 (2012).
% red and blue sidebands,theor

\bibitem{Silveri} M. P. Silveri, J. A. Tuorila, E. V. Thuneberg, and G. S. Paraoanu, Rep. Prog. Phys. \textbf{80}, 056002  (2017).

\bibitem{palermo} A. V. Dodonov, B. Militello, A. Napoli, and A. Messina, Phys. Rev. A \textbf{93}, 052505 (2016).

\bibitem{red1} F. Beaudoin, J. M. Gambetta and A. Blais, Phys. Rev. A \textbf{84}, 043832 (2011).
% master equation, theor

%\bibitem{thermalmachine}
\bibitem{alicki79} R. Alicki, J. Phys. A: Math. Theor. \textbf{12}, L103 (1979).
%\textit{The quantum open system as a model of the heat engine}

%G. Gemma, M. Michel, and G. Mahler, {\it Quantum Thermodynamics}, Springer (2004).

\bibitem{sandoghdar} J. Hwang, M. Pototschnig, R. Lettow, G. Zumofen, A. Renn, S. Gotzinger, and V. Sandoghdar,
%A single-molecule optical transistor
%|doi:10.1038/nature08134
Nature \textbf{460}, 76 (2009).

\bibitem{tsai2010} O. V. Astafiev, A. A. Abdumalikov, Jr., A. M. Zagoskin, Yu. A. Pashkin, Y. Nakamura, and J. S. Tsai,
%Ultimate On-Chip Quantum Amplifier
Phys. Rev. Lett. \textbf{104}, 183603 (2010).

\bibitem{tsai2011} A. A. Abdumalikov, Jr., O. V. Astafiev, Yu. A. Pashkin, Y. Nakamura, and J. S. Tsai,
% Dynamics of Coherent and Incoherent Emission from an Artificial Atom in a 1D Space
Phys. Rev. Lett. \textbf{107}, 043604 (2011).

\bibitem{dv2012A} D. Valente, S. Portolan, G. Nogues, J. P. Poizat, M. Richard, J. M. G\'erard, M. F. Santos, and A. Auff\`eves,
% {\it Monitoring stimulated emission at the single-photon level in one-dimensional atoms}
%DOI: 10.1103/PhysRevA.85.023811
Phys. Rev. A \textbf{85}, 023811 (2012).

%

\bibitem{RabiCircuitQED} J. Braum\"uller, M. Marthaler, A. Schneider, A. Stehli, H. Rotzinger, M. Weides, and A. V. Ustinov,
Nature Comm. \textbf{8}, 779 (2017).
%doi:10.1038/s41467-017-00894-w
%Analog quantum simulation of the Rabi model in the ultra-strong coupling regime

\bibitem{fn} The exact value of the modulation frequency in this case is $\eta_{\mbox{\it \tiny{ADCE}}}=3\omega-\Omega_0 +2(\delta_--\delta_+)(n-1)-2\alpha(n^2-2n+2)$, for $n\geq3$ and $\alpha=g_0^4/\Delta_- ^3$.

\bibitem{kosloff14} A. Levy and R. Kosloff, Europhys. Lett. \textbf{107}, 20004 (2014).
%The local approach to quantum transport may violate the second law of thermodynamics.

%







\end{thebibliography}
\end{document}